\documentstyle[11pt]{article}
\newfont{\frak}{eufm10}
\renewcommand{\theequation}{\arabic{section}.\arabic{equation}}
\begin{document}
\begin{flushright}  Jan 18, 1996  \\
hepth@xxx/9601096\\  ANL-HEP-PR-95-90 \\ KUCP-87 \\  
Miami TH/2/95 \end{flushright}
{\Large {\bf GEOMETRY AND DUALITY IN SUPERSYMMETRIC $\sigma$-MODELS}\\  

Thomas Curtright$^{\S}$, Tsuneo Uematsu$^\natural$, and 
Cosmas Zachos$^{\P}$\\                                        }

$^{\S}$ Department of Physics, University of Miami,
Box 248046, Coral Gables, Florida 33124, USA\\
\phantom{.} \qquad\qquad{\sl curtright@phyvax.ir.Miami.edu}  

$^{\natural}$ Department of Fundamental Sciences, 
FIHS, Kyoto University, Kyoto 606-01, JAPAN\\ 
\phantom{.} \qquad\qquad{\sl uematsu@yukawa.kyoto-u.ac.jp} 

$^{\P}$ High Energy Physics Division,
Argonne National Laboratory, Argonne, IL 60439-4815, USA \\
\phantom{.} \qquad\qquad{\sl zachos@hep.anl.gov}      

\begin{abstract}
The Supersymmetric Dual Sigma Model (SDSM) is a local field theory introduced 
to be nonlocally equivalent to the Supersymmetric Chiral nonlinear 
$\sigma$-Model (SCM), this dual equivalence being proven by explicit canonical 
transformation in tangent space. This model is here reconstructed in 
superspace and identified as a chiral-entwined supersymmetrization of the Dual 
Sigma Model (DSM).  
This analysis sheds light on the {\em boson-fermion symphysis} 
of the dual transition, and on the new geometry of the DSM. 
\end {abstract}
\noindent \rule{7in}{0.1em}
\vskip 0.6cm

\section{Introduction and Conclusions} 
 
A long-standing broad question in field theory 
involves the equivalence of field theories which may appear very different. 
Historically, physicists have been comfortable with local changes of field 
variables (as in the Higgs mechanism) which correspond to point transformations
in classical mechanics: these automatically preserve the canonical structure of
the theory, i.e.~the Poisson Brackets, and the canonical commutators upon
quantization---all of which may also be addressed equivalently in the
conventional functional integral formalism. 

A subtler issue arises, however, when nonlocal transformations are considered,
which link two local field theories. As a rule, nonabelian (t-)duality
transformations in two dimensions (popular in string culture 
\cite{alvarez-gaume,others,giveon}), which broadly map gradients to curls,
hence waves to solitons, are of this type: equivalence results when these are
{\em canonical}\footnote{ i.e.~preserve canonical commutation relations.
Canonical transformations in quantum mechanics underlie Dirac's path integral
formulation \cite{Dirac33}, and have been discussed extensively in field
theory, e.g.~\cite{ghandour,CurtrightGhandour,kastrup}.}
of the non-trivial type that mixes canonical momenta with field gradients, which
results in nonlocal maps.   Conversely, failure of such transformations to
preserve the canonical structure leads to striking inequivalences in such
theories, (cf.~the PCM, a double limit contraction of the WZWN $\sigma$-model,
in \cite{CurtrightZachos'94,pascos}). 

Unfortunately, so far, there is no systematic theory of such transformations 
in field theory, and most discussions merely abstract and change notations on 
a handful of examples. The only examples available are in two dimensions: 
Sine-Gordon/Thirring \cite{Mandelstam};  CM/DSM, 
\cite{CurtrightZachos'94,pascos};
and, finally, SCM/SDSM \cite{sdsm}. The second system was first introduced via
first-order functional integral manipulations \cite{Fridling} and the 
equivalence was shown to be canonical \cite{CurtrightZachos'94} at the 
classical level, and argued \cite{CurtrightZachos'94} to 
extend to the quantized theory\footnote{Ref \cite{subbotin} has
introduced objections to this identification, based on observation
of the respective effective actions at two loops, supported by \cite{palla}. 
We care to suppose that, with proper appreciation 
of the new underlying geometry, some form of identification may eventually 
go through.}; 
moreover, a credible case has been made for complete 
equivalence of the respective S-matrices of the two theories \cite{balog}.
 
The third system was introduced in \cite{sdsm}
and illustrates some intricacies of the nonlocal canonical equivalence best, 
by dint of symphysis, which is the conjoining in the transition maps of 
fermion bilinears with bosons to yield transformed bosons, and of fermions 
with bosons to yield transformed fermions. 
It was first constructed via a generating functional Ansatz \cite{sdsm} and 
first-order lagrangean quadrature of the type \cite{Fridling},
and {\em not} as the supersymmetrization of the bosonic DSM (which, unlike 
the CM, contains torsion). However, we demonstrate here that the {\em same}
model may also be constructed by judiciously supersymmetrizing the bosonic 
DSM: in the past \cite{swzw,geometrostasis},  we have provided a general 
procedure of supersymmetrizing any given torsionful manifold, such as that of 
the DSM. Thus, a superfield construction of the same model from a different 
starting point sheds new light on the supersymmetry realization at work.

Here, we discuss more explicitly detailed aspects of the SDSM, with special 
emphasis on the special realization of supersymmetry and the superfields 
controlling it, to shed light on the intriguing chiral entwining structure 
at work. We review the bosonic case in Section 2; we then review the 
supersymmetric theories 
including tangent space supersymmetric actions in Section 3; we then summarize 
ref \cite{sdsm} by way of introduction of the SDSM, in Section 4. 
In Section 5, we  proceed to rederive the SDSM via superfield extension of the 
DSM and Fridling-Jevicki-type quadrature in 
superspace, and explain and support special features 
observed before which appeared accidental. 
As a guide to insight of the field theory results, a brief review of canonical 
transformations of classical mechanics from the more modern, 
Poisson-Bracket-based point of view is provided in Appendix A.
Subtle relations between curved manifold and tangent space in the dual theory,
specifying a {\em new} inhomogeneous geometry are illustrated in Appendix B.
Every effort is made to stay consistent to the conventions of ref \cite{sdsm};
we correct a typographical error in the part of the action involving a factor
of 3/8 in the quartic fermion interaction. 

\section{Review of the Bosonic Theories} 
\setcounter{equation}{0}
Recall the standard bosonic nonlinear Chiral Model (CM) on 
O(4)~$\simeq$~O(3)$\times$O(3)$\simeq$SU(2)$\times$SU(2).
In geometrical language, 
\begin{equation} 
{\cal L}_{CM}=\frac 12 g_{ab} \partial _\mu\varphi ^a\partial ^\mu \varphi ^b, 
\end{equation}
where $g_{ab}$ is the metric on
the field manifold (three-sphere).
Explicitly, with group elements parameterized as
$U=\varphi ^0+i\tau ^j\varphi ^j$,      $(j=1,2,3)$, where
$(\varphi ^0)^2+{\varphi }^2=1$,
and ${\varphi }^2\equiv $ $\sum_j(\varphi ^j)^2$, we may resolve
$\varphi ^0=\pm \sqrt{1-{\varphi }^2}$, to obtain
\begin{equation} 
g_{ab}= \delta ^{ab}+\varphi^a\varphi^b/(1-\varphi^2) ,
\end{equation}
and hence
\begin{equation} 
{\cal L}_{CM}=\frac 12\left( \delta ^{ij}+\frac{\varphi ^i\varphi ^j}{1-{%
\varphi }^2}\right) \partial _\mu \varphi ^i\partial ^\mu \varphi ^j=
\frac 12 J_\mu^j J^{\mu ~j},
\end{equation}
where we re-expressed the action in terms of tangent
quantities through the use of either left- or right-invariant vielbeine
(e.g.~\cite{geometrostasis})---either choice yields this Sugawara 
current-current form.
Choosing left-invariant dreibeine gives ``$V+A$'' currents which are
vectors on the tangent space.  In terms of the above explicit coordinates,
\begin{equation} 
U^{-1}\partial_{\mu} U= -i~ \tau^j J_\mu^j,
\end{equation}
\begin{equation} 
J^i_\mu = -v_a^{~i} \partial_\mu \varphi^a, 
\end{equation}
where
\begin{equation} 
v_a^{~j}=\sqrt{1-\varphi^2} ~g_{aj} + \varepsilon^{ajb} \varphi^b~.
\end{equation}
Note that these currents are pure gauge, or curvature-free, such that
\begin{equation} 
 \varepsilon ^{\mu \nu }\left( \partial _\mu J^i_\nu +\varepsilon
^{ijk}J^j_\mu J^k_\nu \right) =0.
\end{equation}

This is canonically equivalent to the Dual $\sigma$-Model \cite{Fridling}  
(DSM)\footnote{This model was also considered in the last of 
refs \cite{alvarez-gaume} and the combination of \cite{balog}.} with torsion 
and a new geometry:
\begin{equation}
\!\!\!\!
{\cal L}_{DSM}=\frac 1{1+4{\Phi }^2} 
\left(\phantom{\frac II}\! {\textstyle \frac 12} 
\left(\delta ^{ij}+4\Phi ^i\Phi ^j\right) \partial _\mu \Phi ^i\partial ^\mu
\Phi ^j-\varepsilon ^{\mu \nu }\varepsilon ^{ijk}
\Phi ^i\partial _\mu \Phi ^j\partial _\nu \Phi^k  \right)~ .
\end{equation}
This new geometry is explained in Appendix B, and, in particular, how the 
tangent- and curved-space indices of the field/coordinates $\Phi^i$ in 
this geometry are indistinguishable.

The generator for a canonical
transformation relating $\varphi $ and $\Phi $
at any fixed time is the tangent space functional \cite{CurtrightZachos'94}  
$F[\Phi,\varphi ]=\int_{-\infty }^\infty dx\;\Phi ^iJ_{i}^{1}[\varphi ]$.  
Specifically,
\begin{equation}
\label{F}F[\Phi ,\varphi ]=\int_{-\infty }^{+\infty }dx~\Phi ^i\;\left(
\sqrt{1-{\varphi }^2}\frac {\stackrel{\leftrightarrow} {\partial}}
 {\partial x}\varphi ^i +\varepsilon
^{ijk}\varphi ^j\frac \partial {\partial x}\varphi ^k\right) .
\end{equation}
Although we originally constructed $F$
in the hamiltonian framework by the indirect reasoning reviewed below,
its structure is also evident within the lagrangean framework as follows.
Treating $J$ as independent
variables in ${\cal L}_{CM}=\frac 12 J_\mu^j J^{\mu ~j}$, 
impose, with ref \cite{Fridling}, the pure gauge condition on the currents 
by adding a Lagrange multiplier term,
${\cal L}_{\lambda}=\Phi^j~\varepsilon^{\mu \nu}
(\partial_\mu J_\nu ^j +\varepsilon^{jkl} J_\mu ^k J_\nu ^l)$. Then,
complete a square for the $J$'s and eliminate them
from the dynamics (integrate them out) in favor of the DSM. But to do this, 
one first writes 
${\cal L}_{\lambda}=\partial_\mu (\Phi^j~\varepsilon^{\mu \nu}~J_\nu ^j)
- \varepsilon^{\mu \nu}~J_\nu ^j ~\partial_\mu \Phi^j~
+\varepsilon^{jkl}~\varepsilon^{\mu \nu}~ \Phi^j J_\mu ^k J_\nu ^l$.
The total divergence term, integrated over a world-sheet with
a constant time boundary, gives our generating functional
relating the CM to the DSM\footnote{Roughly speaking, the Lagrange multiplier
dual field $\Phi$ characterizes the ratio of normal magnitudes of the 
Sugawara lagrangian and the zero-curvature constraint, respectively, in the
function space of currents, upon extremization. Recall constrained 
extremization of a surface $f(x,y)-\lambda g(x,y)$ via the Lagrange 
multiplier $\lambda$. Vanishing 
variation specifies that the constraint surface $g(x,y)=0$ touches 
the surface $z=f(x,y)$ at the contact point $(x_0,y_0,z_0=f(x_0,y_0))$.
The section plane $z=z_0$ intersects the respective surfaces 
at two curves which  are tangent to each other, at the point of 
contact: the normals to these two curves on this plane are 
parallel, the ratio of their magnitudes being $\lambda$.}. 

The conjugate momentum of $\Phi ^i$ specified by the generating functional is 
\begin{eqnarray}
\label{Pi} \!\!\! \!\!\!
\Pi _i=\frac{\delta F[\Phi ,\varphi ]}{\delta \Phi ^i}&=&
\sqrt{1-{\varphi }^2}\;\frac \partial {\partial x}\varphi ^i
-\varphi ^i\frac \partial
{\partial x}\left( \sqrt{1-{\varphi }^2}\right) +\varepsilon
^{ijk}\varphi ^j\frac \partial {\partial x}\varphi ^k = \\
&=&\left( \sqrt{1-{\varphi }^2}\;\delta ^{ij}+\frac{\varphi
^i\varphi ^j}{\sqrt{1-{\varphi }^2}}-\varepsilon ^{ijk}\varphi ^k\right)
\frac \partial {\partial x}\varphi ^j=J_{i}^{1} .\nonumber
\end{eqnarray}
The conjugate
of $\varphi ^i$ is
\begin{eqnarray}
\label{PrelimVarPi}\!\!\!\!\!\!
\varpi _i=-\frac{\delta F[\Phi ,\varphi ]}{\delta \varphi ^i}
&=&\left( \sqrt{1-
{\varphi }^2}\;\delta ^{ij}+\frac{\varphi ^i\varphi ^j}{\sqrt{1-{%
\varphi }^2}}+\varepsilon ^{ijk}\varphi ^k\right) \frac \partial {\partial
x}\Phi ^j \nonumber \\
&+&\left( \frac 2{\sqrt{1-{\varphi }^2}}\left( \varphi
^i\Phi ^j-\Phi ^i\varphi ^j\right) -2\varepsilon ^{ijk}\Phi ^k\right) \frac
\partial {\partial x}\varphi ^j,
\end{eqnarray}

To solve for the fields themselves, e.g.~$\Phi[\varphi]$, their canonical 
momenta may be eliminated through substitution for $\Pi _i$ and $\varpi_i$, 
in terms of 
$\partial\varphi^j/{\partial t}$ and $\partial \Phi^j /{\partial t}$, as 
follows from ${\cal L}_1$ and ${\cal L}_3$:
\begin{equation}
\Pi _i=\frac 1{1+4{\Phi }^2}\left( \left( \delta ^{ij}+4\Phi ^i\Phi
^j\right) \frac \partial {\partial t}\Phi ^j+2\varepsilon ^{ijk}\Phi ^j\frac
\partial {\partial x}\Phi ^k\right) ,~~~~~~~~
\varpi _i=\left( \delta ^{ij}+%
\frac{\varphi ^i\varphi ^j}{1-{\varphi }^2}\right) \frac \partial
{\partial t}\varphi ^j.
\end{equation}

However, the resulting transformation laws are complicated and nonlocal,
as illustrated at the end of this section. Instead, it is relatively more 
instructive to simply identify the conserved, curvature-free current in the 
two theories, 
consistently to the above, an identification which will turn out to be local. 
It is then straightforward to exploit the current-current form of the
respective hamiltonian densities which will thus likewise identify. 

Now, then, in the DSM, what is the conserved, curvature-free current? In
contrast to the PCM, where it was essentially {\em forced} to be a topological
current, here a topological current by itself will not suffice; neither will
a conserved Noether current. (Under isospin transformations, 
$\delta \Phi ^i=\varepsilon ^{ijk}\Phi ^j\omega ^k$, the Noether
current of ${\cal L}_3$ is $I_i^\mu =\delta {\cal L}_3/\delta 
(\partial _\mu \omega ^i)$ so $I_i^0=\varepsilon ^{ijk}\Phi
 ^j\Pi _k$, but it is not curvature-free.)
Instead, the conserved,  curvature-free current
 ${\cal J}_i^\mu [\Phi ,\Pi ] =J_i^\mu [\varphi ,\varpi ]$
(identified with $J_i^\mu $ of  the CM) is a {\em mixture} of the Noether
isocurrent and a topological current:
${\cal J}_i^\mu =2I_i^\mu -\varepsilon ^{\mu \nu }\partial _\nu \Phi ^i$, so
that ${\cal J}_i^1=\Pi _i$. Both conservation and curvature-freedom now hold
on-shell, for the on-shell identified current\footnote{The equations of motion
are necessary, as Hamilton's equations have already been utilized in the
elimination of the canonical momenta from the subsequent expressions.}:
\begin{equation}
\label{PhiCurrent}{\cal J}_i^\mu =\frac{-1}{1+4{\Phi }^2}\biggl( \left(
\delta ^{ij}+4\Phi ^i\Phi ^j\right) \varepsilon ^{\mu \nu }\partial _\nu
\Phi ^j+2\varepsilon ^{ijk}\Phi ^j\partial ^\mu \Phi ^k\biggr) .
\end{equation}
Nevertheless, the following canonical identifications of currents 
can be shown \cite{CurtrightZachos'94} to hold off-shell:
\begin{equation}
\label{Jspace}{\cal J}_i^1\equiv \Pi _i=\left( \sqrt{1-{\varphi }^2}%
\;\delta ^{ij}+\frac{\varphi ^i\varphi ^j}{\sqrt{1-{\varphi }^2}}%
-\varepsilon ^{ijk}\varphi ^k\right) \frac \partial {\partial x}\varphi
^j\equiv J_i^1,
\end{equation}
\begin{equation}
\label{Jtime}{\cal J}_i^0\equiv -\frac \partial {\partial x}\Phi
^i-2\varepsilon ^{ijk}\Phi ^j\Pi _k=-\sqrt{1-{\varphi }^2}\;\varpi
_i-\varepsilon ^{ijk}\varphi ^j\varpi _k\equiv J_i^0.
\end{equation}
These two relations combine to integrate to the explicit nonlocal map 
\cite{pascos},
\begin{equation} 
\Phi (x)\cdot\tau =  U^{-1}(x) U(0)\Phi (0) \cdot\tau U^{-1}(0)U(x)+
U^{-1}(x) \Bigl(\int_0^x \! dy ~i U(y)    \partial_0 U^{-1}(y) \Bigr)  U(x)~~.
\end{equation}
The dual character of this nonlocal transition is manifest in the weak field 
limit. It bears repeating that canonical commutators of such complicated 
nonlocal expressions are simple and conventional, precisely because the 
transformation is canonical: e.g.~equal time commutators of two expressions of 
this type at different space points $x$ and $z$ vanish, as these expressions 
are essentially local, despite formidable appearances! 

Substituted into the Sugawara-current-current hamiltonians, these locally 
identified currents (\ref{Jspace},\ref{Jtime}) further lead to mutual local 
identification of the respective hamiltonian densities, 
\begin{equation}
{\cal H}_{CM}=J_0 J_0 + J_1 J_1={\cal J}_0 {\cal J}_0 + 
{\cal J}_1 {\cal J}_1 ={\cal H}_{DSM}  ~.
\end{equation}

In terms of the dreibeine of the new dual geometry (discussed in Appendix B),
the currents of the DSM read 
\begin{equation}
{\cal J}_j^\mu=-(V_{(ja)} \varepsilon^{\mu\nu} \partial_\nu \Phi^a +V_{[ja]}
\partial^\mu \Phi^a).  \label{dualcurrent}
\end{equation}
The reader may contrast these currents with those of a WZWN model on a group 
manifold. The currents for the latter are obtained by keeping the vielbein 
intact: ${J}%
_\mu ^j=-{v}_a^{\;j}\left( \partial _\mu \phi ^a+\eta \varepsilon _{\mu \nu
}\partial ^\nu \phi ^a\right) $. (For the CM, $\eta=0$). Also note that the DSM
dreibein already appeared in the above action, 
${\cal L}_{DSM} ={\textstyle \frac 12}
V_a^j \left( \partial _\mu \Phi ^a\partial ^\mu \Phi ^j +\varepsilon ^{\mu
\nu } \partial _\mu \Phi ^a \partial _\nu \Phi ^j \right) $.

\section{Supersymmetric Theories}
\setcounter{equation}{0}
In a direct application of the general construction for supersymmetric 
$\sigma$-models with torsion \cite{geometrostasis,swzw} (whose conventions
we use), the supersymmetric extensions of the two bosonic models
above through the addition of Majorana fermions, the SCM \cite{witten} without 
torsion, and the $sdsm$  with torsion, are readily read off. 
In following sections, we canonically transform between the two.  
We first review this in component formalism \cite{sdsm}  but in the 
next section we will provide a complementary picture in superspace which 
illuminates and confirms our construction. 

The SCM is 
\begin{eqnarray}
{\cal L}_{SCM}&=&\frac 12\left( g_{ab}~\partial _\mu \varphi ^a
\partial ^\mu \varphi ^b+ i g_{ab}\overline{\psi}^a  /\kern-.7em D \psi^b+ 
\frac 16 R_{abcd} \overline{\psi}^a\psi^c\overline{\psi}^b\psi^d\right)\\
&=&\frac 12\left( g_{ab}~\partial _\mu \varphi ^a
\partial ^\mu \varphi ^b+ i g_{ab}\overline{\psi}^a  /\kern-.7em D \psi^b+ 
\frac 14 (g_{ab} \overline{\psi}^a\psi^b)^2 \right)~,\nonumber
\end{eqnarray}
where 
\begin{equation}
D_\mu \psi^b= \partial_\mu \psi^b+ \Gamma^b_{cd}\partial_\mu \varphi^c 
\psi^d  ,  \qquad \Gamma^a_{bc}=\varphi^a g_{bc}=
v^{aj}\partial _{(b}v_{c)}^{~~j}, 
\qquad R_{abcd}= g_{ac}g_{bd}-g_{ad}g_{bc}=R_{cdab} ~.
\end{equation}
\begin{equation}
g^{ab}=\delta^{ab} - \varphi^a \varphi^b=v^{aj}v^{bj}, \qquad
v^{aj}= \sqrt{1-\varphi^2} ~\delta^{aj} + \varepsilon^{ajb} \varphi^b,
\qquad v_a^{~j}=\sqrt{1-\varphi^2} ~g_{aj} + \varepsilon^{ajb} \varphi^b~.
\end{equation}

The Cartan-Maurer relations of the Dreibeine merit recall:
\begin{equation}
\frac 12 ( \partial_a v_b^{~j}-\partial_b v_a^{~j})=
\varepsilon^{jkl} v_a^{~k} v_b^{~l} = 
\varepsilon_{jab} 
+{\delta^{ja} \varphi^b - \delta^{jb} \varphi^a \over 
\sqrt{1-{\varphi}^2}}~.
\end{equation}

The conserved currents now consist of the previous bosonic terms
augmented by spinor bilinears.  
\begin{equation}
C^i_\mu = J^i_\mu+K^i_\mu~,~~~~~
J^i_\mu = -v_a^{~i} \partial_\mu \varphi^a~,~~~~~
K^i_\mu = \frac i2~\varepsilon^{ijk} v_a^{~j} v_b^{~k} 
\overline{\psi}^a \gamma_\mu \psi^b~.
\end{equation}
The last term, explicitly, is given by the above results 
for the Cartan-Maurer relations,
\begin{equation}
K^i_\mu = \frac 12~i~\varepsilon^{iab} \overline{\psi}^a \gamma_\mu \psi^b
-i~ {\varphi^a \overline{\psi}^a \gamma_\mu \psi^i 
\over \sqrt{1-{\varphi}^2}}~.
\end{equation}

This is to be expected: The tangent-space left-invariant spinor of (A.51) 
specified in ref \cite{geometrostasis}, 
\begin{equation}  
\chi^j= v_a^j \psi^a ,
\end{equation}
transforms as $\delta \chi^j= \varepsilon^{jkl}\xi^k\chi^l$ under a full 
$V+A$ transformation. The right rotation in tangent space transforms 
linearly, and the spinor's contribution to the corresponding current is that of 
a conventional isorotation. Recall (ref \cite{geometrostasis}, (A.52), (A.29))
that now the lagrangean  simplifies significantly, to a mere function of 
tangent-space spinors and currents:  
\begin{equation}   
{\cal L}_{SCM}=\frac 12 \biggl( J_\mu^j J^{\mu ~j}+ 
i \overline{\chi }^j /\kern-.5em\partial   ~\chi^j
+i\varepsilon^{jkl}\overline{\chi}^j /\kern-.5em J^k \chi ^l+
\frac 14 (\bar \chi \chi )^2 \biggr).
\end{equation}  
This tangent space formulation is at the heart of the 
canonical transformation, as will become apparent in the following section. 

The supersymmetry transformation laws are
\begin{equation}   
\label{susytfmation}
\delta \chi^j= i/\kern-.5em J^j \epsilon -\frac 12 \varepsilon^{jkl} 
(\gamma_p \epsilon ~\bar \chi^k \gamma_p \chi^l + 
\gamma_\mu \epsilon ~ \bar \chi^k \gamma^\mu \chi^l ) ,\qquad 
\delta J_\mu^j=-\bar\epsilon(\partial_\mu \chi^j + \varepsilon^{jkl} J_\mu 
^k\chi^l)~.
\end{equation}

 The bosonic generating functional can then be re-presented as 
\begin{equation}
F[\Phi,\varphi ]
=\int_{-\infty }^\infty dx\;\Phi^i v^i_a \partial_1 \varphi^a~.
\end{equation}

The above supersymmetry transformations follow from the general case:
\begin{equation}
\delta \varphi = \bar \epsilon \psi \quad , \qquad
\delta \psi= (\hbox{{\frak F}} -i~/\kern-.5em\partial  \varphi)~
\epsilon\quad , \qquad
\delta \hbox{{\frak F}} =-i\bar \epsilon /\kern-.5em\partial \psi \quad ,
\end{equation}
where we use the conventions of ref \cite{geometrostasis} eq (5.7),
\begin{equation}
2\hbox{{\frak F}}^a=  \Gamma^a_{bc} \overline{\psi}^b \psi^c -S^a_{bc}
\overline{\psi}^b \gamma_p \psi^c ,
\end{equation}
where the torsion $S$ vanishes here---but, of course, {\em not} for the 
${\cal L}_{sdsm}$ model, below. 

In this case, the SCM equations of motion, written covariantly, directly 
lead to conservation of the supercurrent of ref \cite{witten}:
\begin{equation}   \label{supcurr} 
s_\mu =-i/\kern-.5em \partial \varphi^a \gamma_\mu g_{ab} \psi^b~
= i /\kern-.5em J^j \gamma_\mu \chi^j ~.
\end{equation}

Correspondingly, we may express the lagrangean in terms of conserved vector
currents.
\begin{equation}   
{\cal L}_{SCM}=\frac 12 \left( C_\mu^j C^{\mu ~j}+ 
i \overline{\chi }^j /\kern-.5em\partial   ~\chi^j
+\frac 34 (\bar \chi^j \chi^j )^2 \right),
\end{equation}  
with supersymmetry transformations 
\begin{equation}      
\delta \chi^j= i/\kern-.5em C^j \epsilon -{\textstyle\frac 12} 
\gamma_p \epsilon \varepsilon^{jkl}  (~\bar \chi^k \gamma_p \chi^l),\qquad 
\delta C_\mu^j=-\bar\epsilon(\partial_\mu \chi^j 
+ \varepsilon^{jkl} \varepsilon_{\mu}^{~ \nu} C_{\nu}^{~k} \gamma_p \chi^l)~.
\end{equation}

Taking $n=3/2$ in ref \cite{geometrostasis}, the conventional 
supersymmetrization ${\cal L}_{sdsm}$ of the DSM (which, unlike the SCM, 
{\em contains} torsion) is readily seen to be  
\begin{equation}
{\cal L}_{sdsm} =\frac 12\left( G_{ab}~\partial _\mu \Phi^a\partial ^\mu \Phi^b
+ i G_{ab}\overline{\Psi}^a  /\kern-.7em {\cal D} \Psi^b+
E_{ab}\varepsilon^{\mu\nu} \partial_\mu \Phi^a \partial_\nu \Phi^b +
\frac 18 {\cal R}_{abcd} \overline{\Psi}^a(1+\gamma_p)  \Psi^c 
\overline{\Psi}^b(1+\gamma_p)\Psi^d\right),
\end{equation}
with a {\em new geometry} elaborated in Appendix B. The curly-covariant 
derivative on the fermions is
\begin{equation}
{\cal D}_{\mu} \Psi^a= \partial_\mu \Psi^a + \Gamma^a_{bc} \Psi^b
\partial_\mu \Phi^c - S^a_{bc} \Psi^b \varepsilon_{\mu\nu} \partial^\nu
\Phi^c ,
\end{equation}
in terms of the tensors of the new geometry specified in Appendix B. 

\section{Canonical Equivalence of the Supersymmetric Models}
\setcounter{equation}{0}
If these two supersymmetric theories are canonically equivalent like their
bosonic limits, a generating functional in tangent space for such a 
canonical transformation 
is needed. Taking into consideration dimensional consistency, 
Lorentz-invariance, and a good free-field limit,
we posit the Ansatz for the supersymmetric theory
in tangent space relating 
$\varphi $ and $\chi$ at any fixed time to $\Phi $ and $X$ 
(the bosons and fermions of the dual theory):
\begin{equation}
\label{FS} F[\Phi,X ,\varphi,\chi] =
\int \!dx\;
\left(\Phi^j J^{1~j}[\varphi ]- \frac i2~\overline X^j \gamma^1 \chi^j \right)
=\int dx\Bigl(\Phi ^i (
\sqrt{1-{\varphi }^2}\frac {\stackrel{\leftrightarrow} {\partial}}
 {\partial x}\varphi ^i +\varepsilon
^{ijk}\varphi ^j\frac \partial {\partial x}\varphi ^k) -\frac i2
\overline X^j \gamma^1 \chi^j\Bigr)~.
\end{equation}

Classically, the canonical conjugate to $\chi$, 
\begin{equation}                  
\pi_{\chi} \equiv \delta {\cal L}_{SCM}/\delta \partial_0 \chi 
= -i \chi^{\dag}/2
\end{equation} 
is obtained from $F$ as  
\begin{equation}                  
-\delta F/\delta \chi = -iX^{\dag} \gamma_p /2 , 
\end{equation}
where 
$\gamma_p\equiv\gamma^0 \gamma^1$. So under the canonical transformation
\begin{equation}
\label{chiX}
\chi^j=\gamma_p X^j .
\end{equation}

Likewise, the momentum conjugate to $X$ is 
\begin{equation}                  
\delta F/\delta X =
- i \chi^{\dag} \gamma_p/2,
\end{equation} 
leading to 
\begin{equation}                  
\pi_{_X} \equiv -i X^{\dag}/2, 
\end{equation}
which specifies part of the dual lagrangean.

This chiral rotation of the fermions reflects the duality transition 
of their bosonic superpartners, whose gradients map to curls 
in the weak field limit, as already noted. (For a mathematically subtle  
interpretation of tangent space canonical transformation in terms of 
Cartan's equivalence problem see ref \cite{alvarezliu}.) 

As a result of (\ref{chiX}), the equal-time anticommutation 
relations for Majorana spinors in tangent space, 
\begin{equation}                  
\{  \chi^j(x), \chi^k (y)\}=
\{  X^j(x), X^k (y)\}=2~ \delta^{jk} \delta (x-y)~,
\end{equation} 
are preserved in the above transformation, thus confirming its 
identification as canonical. 

To handle the bosons, it may be advantageous to resort to 
the first order Lagrangean quadrature mentioned before. 
Adding the customary pure-gauge-enforcing Lagrange multiplier to 
the tangent space SCM and integrating by parts leads to 
\begin{equation}
{\cal L}_{SCM\lambda}=\frac 12 \left( J_\mu^j M^{\mu \nu}_{jk} J_{\nu}^k 
-2 \varepsilon^{\mu \nu}~J_\nu ^j ~\partial_\mu \Phi^j~+
i \overline{\chi }^j /\kern-.5em\partial   ~\chi^j
+i\varepsilon^{jkl}\overline{\chi}^j /\kern-.5em J^k \chi ^l+
\frac 14 (\bar \chi \chi )^2 \right), 
\end{equation}  
where 
\begin{equation}
M^{\mu \nu}_{ab}\equiv g^{\mu \nu}~ \delta_{ab}
+2\varepsilon^{\mu \nu}~\varepsilon^{abc}\Phi^c ,
\end{equation}
with an inverse which satisfies $M^{\mu \nu}_{ab} (N_\nu^{~\lambda})_{bc} 
= g^{\mu \lambda} \delta_{ac}$: 
\begin{equation}
N^{\mu \nu}_{ab} = g^{\mu \nu}~ G_{ab}+\varepsilon^{\mu \nu}~E_{ab} .
\end{equation}

The crucial transition bridge to the SDSM relies on the bosonic current 
encountered in section 2, 
\begin{equation}                  
{\cal J}_j^\mu = -N^{\mu \nu}_{jk} 
\varepsilon_{\nu\lambda}\partial^\lambda \Phi^k =
  \frac{-1}{1+4\Phi^2}\Bigl( (\delta ^{jl}+4\Phi^j\Phi^l) 
\varepsilon ^{\mu \nu }\partial _\nu
\Phi ^l+2\varepsilon ^{jlk}\Phi ^l\partial ^\mu \Phi ^k\Bigr) ,
\end{equation}
which conjoins  with the fermionic bilinear component into 
\begin{equation}                  
\label{symphysis}
J^{\mu~j}=-N^{\mu \nu}_{jb} 
\bigl( \varepsilon_{\nu\lambda}\partial^\lambda \Phi^b
-\frac i2 \varepsilon^{bcd} \overline{\chi}^c \gamma_\nu \chi^d \bigr)
={\cal J}^{\mu j} + N^{\mu\nu}_{jk} K^k_{\nu}, 
\end{equation}
which follows from varying $J$ in the first order lagrangean 
${\cal L}_{SCM\lambda}$.
(Note from (\ref{chiX}), $K_\mu ^j [\chi] =K_\mu ^j [X]$.)
This is the on-shell relation linking $J$ to $\Phi$. Below, this is shown to be 
derivable from the generating functional $F_s$. 

The lagrangean resulting from Fridling-Jevicki-type 
\cite{Fridling} substitution for $J$, or,
equivalently, completing the quadratic in ${\cal L}_{SCM\lambda}$ and
integrating the shifted $J$s out, is \cite{sdsm}:
\begin{eqnarray}
{\cal L}_{SDSM}&=&  
\frac 12 i \overline{\chi }^j /\kern-.5em\partial   ~\chi^j
+\frac 18 (\bar \chi \chi )^2 
-\frac 12 \bigl( \varepsilon_{\mu\lambda}\partial^\lambda \Phi^a
-\frac i2 \varepsilon^{acd} \overline{\chi}^c \gamma_\mu \chi^d \bigr)
N^{\mu \nu}_{ab} \bigl( \varepsilon_{\nu\rho}\partial^\rho \Phi^b
-\frac i2 \varepsilon^{bef} \overline{\chi}^e \gamma_\nu \chi^f \bigr)
\nonumber \\  
&=&  \frac i2 \overline{X }^j /\kern-.5em\partial   ~X^j
+\frac i2 \varepsilon^{ijk} \overline {X}^i/\kern-.7em{\cal J}^j X^k
+(\overline X^i X^i ) \left(\frac 38  (\overline X^j X^j ) +
  { (\varepsilon^{jkl} \overline X^j \gamma_p X^k \Phi^l
-2 \overline X^j X^k  \Phi^j \Phi^k )\over (1+4\Phi^2)} \right)
\nonumber \\
& &+\frac 12\left( G_{jk}~\partial _\mu \Phi^j\partial ^\mu \Phi^k
+ E_{jk}\varepsilon^{\mu\nu} \partial_\mu \Phi^j \partial_\nu \Phi^k \right).
\end{eqnarray}
This result forces the
variation $\delta / \delta \Phi^a$ of the generating functional
to be just the $\varphi$ bosonic current $J$, while
the $\chi^j=\gamma_p X^j$'s are regarded as independent variables.
As a result, the generating functional automatically yields 
\begin{equation}
\Pi= J^1={\cal J}^1 + (N\cdot K)^1.
\end{equation}

We now use the variation w.r.t. $\varphi$ to match 
the timelike components as well. The arguments of the 
bosonic model ref \cite{CurtrightZachos'94}  
eqs (3.11, 3.12), which connect timelike components of currents) 
remain exactly as they were,
\begin{equation}
\label{boscur}  -\frac \partial {\partial x}\Phi
^i-2\varepsilon ^{ijk}\Phi ^j\Pi _k=-\sqrt{1-{\varphi }^2}\;\varpi
_i-\varepsilon ^{ijk}\varphi ^j\varpi _k 
\end{equation}
since both sides are equal to 
$$
-\frac \partial {\partial x}\Phi ^i+2\Phi ^j\left( \varphi
^j\frac \partial {\partial x}\varphi ^i-\varphi ^i\frac \partial {\partial
x}\varphi ^j\right) +2\varepsilon ^{ijk}\Phi ^j\left( \varphi ^k\frac
\partial {\partial x}\left( \sqrt{1-{\varphi }^2}\right) -\sqrt{1-{%
\varphi }^2}\frac \partial {\partial x}\varphi ^k\right); 
$$
but now $\Pi$ contains an additional fermionic piece beyond its bosonic 
component, as seen above. And likewise $\varpi$:
\begin{equation}
\varpi^i= bosonic~~~+ K_0^j (\sqrt{1-\varphi^2} \delta^{ij} 
+ \varphi^i \varphi^j /\sqrt{1-\varphi^2} +
\varepsilon^{ijl}\varphi^l )~.
\end{equation}
These then introduce fermionic current pieces in the above eq (\ref{boscur}), 
\begin{equation} 
{\cal J}_0^i - 2 \varepsilon^{ijk}\Phi^j (N\cdot K)^{k1} = J^i_0 -K_0^i
\end{equation}
As a direct consequence, 
\begin{equation}         
J^0={\cal J}^0 + (N\cdot K)^0, 
\end{equation}
and the above pivotal identification of currents (\ref{symphysis}) holds. 
Note how duality is implemented on these manifolds: 
{\em Bosons in one theory contain both bosons and fermions in the dual
theory.} 

Further note that the conserved current is 
\begin{equation}         
{\cal J}^{\mu j} + N^{\mu\nu}_{jk} K^k_{\nu}+K^{\mu j}. 
\end{equation}         

The supercurrent for the DSM is 
\begin{equation}   \hbox{{\frak S}}_\mu= i\gamma_p (
/\kern-.6em{\cal J}^j + /\kern-.7em N\cdot K^j 
)\gamma_\mu X^j, \end{equation}
which holds by the symphysis identification of currents (\ref{symphysis}) above. 
Consequently, supercharges identify, and whence hamiltonians (which are their
squares), energy-momentum tensors, and all such quantities conventionally
constrained by supersymmetry. 

From the structure of this supercurrent, it then follows that 
this is a ``chirally twisted" realization of supersymmetry: 
\begin{equation}
\delta \Phi^a= \overline \epsilon 
(X^a +2\gamma_p \varepsilon^{abk} \Phi^b X^k) ,
\end{equation}
which {\em preserves the parity of the original} $X$, and thus of 
the action. This is the combination
entering into the spinors of the ${\cal L}_{sdsm}$ model (the fermion 
entering in the 
respective superfield of the next section) for comparison with the 
SDSM:
\begin{equation}
\Psi^a = X^a +2\gamma_p \varepsilon^{abj} \Phi^b X^j
= (V^{(aj)} + \gamma_p V^{[aj]} ) X^j. 
\end{equation}
\begin{equation}
(1+ \gamma_p ) \Psi^a  = (1+\gamma_p) V^{aj} X^j,
\end{equation}
\begin{equation}
(1- \gamma_p ) \Psi^a  = (1-\gamma_p) \tilde V^{aj} X^j,
\end{equation}
where $\tilde V^{ai}$ is the transpose of $V^{ai}$. Consequently,
its inverse $\tilde V_{ai}=G_{ai}-E_{ai}$ is also the transpose of 
$V_{ai}$. As a result, 
\begin{equation}
X_j = (V_{(aj)} + \gamma_p V_{[aj]} ) ~\Psi^a.\label{bridge} 
\end{equation}
This last relation strongly echoes eq (\ref{dualcurrent}), which finds its 
explanation in the superfield formulation below. 

It turns out that ${\cal L}_{SDSM}= {\cal L}_{sdsm}$. 
The pure bosonic piece of ${\cal L}_{SDSM}$ above is, naturally, 
${\cal L}_{DSM}$.
To further match with the pieces of ${\cal L}_{sdsm}$ quadratic and quartic 
in fermion fields, respectively, we need check the following two relations.

For the function $W$ in ${i\over 2} \overline X^i /\kern-.5em\partial 
\Phi^m W_{ikm} X^k$, one needs show 
\begin{equation} 
\varepsilon^{ikj} G_{jm} \gamma_p +  \varepsilon^{ikj} E_{jm} =
2  (-\delta^{ai}\gamma_p +2 \varepsilon^{ain}\Phi^n) G_{ab} \varepsilon^{bkm} 
+(\delta^{ai}-2\gamma_p \varepsilon^{ain}\Phi^n) (\Gamma_{abm}-\gamma_p S_{abm})
(\delta^{bk}-2 \gamma_p \varepsilon^{bks}\Phi^s), 
\end{equation}
which, indeed, holds. 

Secondly, for the fermion quartic terms, use is made of the group-Fierz 
identities
\begin{equation}
\bigl( \Phi \cdot \overline{X}~X\cdot \Phi \bigr) \bigl( \varepsilon
^{jkl}\Phi ^l\overline{X}^j\gamma _pX^k\bigr) ~=\bigl( \Phi ^2~\overline{X}%
\cdot X\bigr) \bigl( \varepsilon ^{jkl}\Phi ^l\overline{X}^j\gamma _pX^k%
\bigr) =\frac 12\,\Phi ^2\,\varepsilon ^{jkl}\,\bigl( \Phi \cdot \overline{X}%
~X^l\bigr) \bigl( \overline{X}^j\gamma _pX^k\bigr) \;,
\end{equation}
to prove 
$$
\bigl(\overline X^i X^i \bigr)  
\Bigl(\frac38   \bigl(\overline X^j X^j \bigr) 
-\frac{2}{(1+4\Phi^2)} \bigl( \Phi^j \overline X^j \Phi^k X^k \bigr)
+\frac{1}{(1+4\Phi^2)} \bigl( \varepsilon^{jkl} \Phi^l 
\overline X^j \gamma_p X^k \bigr) \Bigr)=
$$ 
\begin{equation}
=\frac 1{16} {\cal R}_{abcd} \bar\Psi^a(1+\gamma_p)\Psi^c 
\bar\Psi^b(1+\gamma_p)\Psi^d= \frac 1{16} {\cal R}_{abcd} \tilde V^a_i V^c_j 
\tilde V^b_k V^d_l \bar X^i(1+\gamma_p)X^j\bar X^k(1+\gamma_p)X^l~.
\end{equation}
Consequently, supersymmetrization of the mutually dual bosonic models
produces mutually dual theories, as demonstrated:
\def\mapright#1{\vbox{\ialign{##\crcr $\hfil\scriptstyle{\ #1 \ }\hfil$   
    \crcr\noalign{\kern+1pt\nointerlineskip}\rightarrowfill \crcr} }}
\def\mapdownl#1{\lower1.4ex\hbox{\llap{$\vcenter{\hbox{$\scriptstyle#1$}}$}}
     \lower1.4ex\hbox{\Big\downarrow}}
\def\mapdownr#1{\lower1.4ex\hbox{\Big\downarrow} \lower1.4ex\hbox{
    \rlap{$\vcenter{\hbox{$\scriptstyle#1$}}$}}}
\begin{equation} 
\matrix{    CM  & \mapright{~susy~~} & SCM \cr 
\mapdownl{dual} & \  & \mapdownr{dual} \cr
DSM & \mapright{~susy~~} sdsm &=SDSM.  \cr                }
\end{equation}

In these variables, the supercurrent for the DSM now reads,
\begin{equation}
\hbox{{\frak S}}_\mu =i/\!\!\!\partial \Phi ^a\gamma _\mu G_{ab}\Psi
^b+i\gamma _p\gamma ^\nu \gamma _\mu (N\cdot K)_\nu ^{~j}~(V_{(aj)}+\gamma
_pV_{[aj]})\Psi ^a.
\end{equation}
which may now be compared to eqn (\ref{supcurr}). 

\section{Superspace Formulation}
\setcounter{equation}{0}
General superfield formulations for supersymmetric $\sigma$-models with
torsion are given in ref \cite{geometrostasis}. 
Recall 
\begin{equation}  
D={ \partial\over \partial\overline{\theta}} -i\partial\!\!\!/ \theta ,
\qquad  \qquad \overline{D}=-{\partial\over \partial\theta} +i \bar{\theta} 
\partial\!\!\!/ ,
\qquad  \qquad Q={ \partial\over  \partial\overline{\theta}} +i\partial\!\!\!/ 
\theta  ~.
\end{equation}

The scalar superfield for the supersymmetric chiral model, SCM, is
\begin{equation}   
\hbox{{\boldmath $\varphi$}}^a=\varphi ^a+\overline{\theta }\psi^a
+\frac 12\overline{\theta }\theta ~ \hbox{{\frak F}}^a
\end{equation}
and has superderivatives
\begin{eqnarray}
D\hbox{{\boldmath $ \varphi$\unboldmath}}^a
&=&\psi ^a+\left(\hbox{{\frak F}}^a
-i\partial \!\!\!/\varphi^a\right) \theta 
+\frac i2\,\partial \!\!\!/\,\psi ^a\overline{\theta }\theta \,\nonumber \\
\overline{D}\hbox{{\boldmath $ \varphi$\unboldmath }}^a
&=&\overline{\psi }^a+\overline{\theta }\left(\hbox{{\frak  F}}^a
+i\partial \!\!\!/\varphi ^a\right) -\frac
i2\,\overline{\theta }\theta \,\overline{\psi }^a  
{\stackrel{\leftarrow} {\partial}}\!\!\!\!/~.
\end{eqnarray}
From this, one may construct the bilinear
\begin{eqnarray}
\overline{D}\hbox{{\boldmath $ \varphi$}}^a\gamma _p 
D{\hbox {\boldmath $\varphi$}}^b &=&\overline{\psi }^a\gamma _p \psi ^b+%
\overline{\psi }^a\gamma _p \left(\hbox{{\frak  F}}^b
-i\partial \!\!\!/\varphi^b\right) \theta 
-\overline{\psi }^b\gamma _p\left(
\hbox{{\frak F}}^a-i
\partial\!\!\!/ \varphi ^a\right) \theta \nonumber \\
&+&\,\partial _\mu \left[ \frac i2\,\overline{\psi }^a
\gamma _{\nu }\psi ^b-\varphi ^a\partial _\nu \varphi ^b\right] 
\varepsilon ^{\mu\nu }\,\overline{\theta }\theta .
\end{eqnarray}

The corresponding superfield for the dual theory is 
\begin{equation}   
{ \bf \Phi }^a =\Phi ^a+\overline{\theta }\Psi 
^a+\frac 12\overline{\theta }\theta \,Y^a,
\end{equation}
hence
\begin{equation}
D { \bf \Phi }^a =\Psi ^a+\left(Y^a
-i\partial \!\!\!/\Phi^a\right) \theta 
+\frac i2\,\partial \!\!\!/\,\Psi ^a\overline{\theta }\theta . 
\end{equation}

The dual transition between two field theories, however, is normally 
effected in tangent space.
To address tangent space in superspace, start from the chiral element superfield
\begin{equation}
{ \bf  G} = U(x)\left(\hbox{{1}\kern-.25em\hbox{l}}
+i~\overline{\theta}\chi\cdot\tau +   
{\overline{\theta}\theta \over 2} (i Z\cdot \tau + {\overline{\chi}\chi\over 2}
)\right)~,
\end{equation}
a most general Ansatz, such that, for unconstrained $\chi$ and $Z$,
\begin{equation}  
{\bf  G}^\dagger~ {\bf  G} =\hbox{{1}\kern-.25em\hbox{l}}~.
\end{equation}

Recalling~  $U^{-1} \partial_{\mu} U = -i \tau\cdot J_{\mu}$, obtain its 
superspace analog \cite{schonfeld}, the spinorial {\em current superfield}, 
$$
{\bf  G}^{-1} D {\bf  G} = i \chi\cdot \tau +
 (i Z\cdot \tau - J\!\!\!\!/ \cdot \tau) \theta
 -{i\over 2}  \gamma_{\mu}\theta \varepsilon^{jkl}\tau^j\overline{\chi}^k
\gamma^\mu \chi^l 
 -{i\over 2}  \gamma_p\theta \varepsilon^{jkl}\tau^j\overline{\chi}^k
\gamma_p \chi^l +
$$
\begin{equation} 
+ {\overline{\theta} \theta\over 2}(2i \varepsilon^{jkl} \tau^jZ^k\chi^l-
\partial\!\!\!/\chi\cdot \tau -2 \varepsilon^{jkl}\tau^j 
{J^k}\!\!\!\!\!\!/ \ \chi^l + i \chi\cdot\tau ~\overline{\chi}\cdot\chi)~, 
\end{equation}
notably valued in the SU(2) Lie algebra (unlike {\bf G}). 
Consequently, the tangent space SCM action is specified in superspace by 
\begin{equation}
{\cal L}_{sfield} = 
\frac 1 8 \int d^2 \theta~ \hbox{Tr} \overline{D{\bf G}}D{\bf G}=
\frac 1 2 \left( Z^2 + J_{\mu}\cdot J^{\mu}+
 i\overline{\chi}\cdot \partial\!\!\!/ \chi + 
i \varepsilon^{jkl}\overline{ \chi}^j J\!\!\!\!/^k\chi^l+
\frac 1 4 (\overline{\chi}\cdot\chi)^2\right).   \label{SCMsuperfield}
\end{equation}
Thus, $Z=0$ on-shell, matching the component supersymmetry 
transformations on the tangent space objects of the SCM exhibited in eq 
(\ref{susytfmation}).
The variational result for the (right-) conservation law in superspace is thus 
\begin{equation}
\int d^2 \theta~ \overline{ D} ({\bf G}^{-1}D{\bf G})=-2i
\partial_{\mu}  \left( J^{\mu}\cdot \tau + {i\over 2} 
\varepsilon^{jkl}\tau^j \overline{ \chi}^k \gamma^{\mu} \chi^l\right), 
\end{equation}
which again identifies with the conserved current.

Now consider the superspace curvature identity \cite{schonfeld}
\begin{equation}
\overline{ D}\gamma_p ({\bf G}^{-1}D{\bf G})  + ({\bf G}^{-1}\overline{
D}{\bf G})  \gamma_p ({\bf G}^{-1}D{\bf G}) =0~. 
\end{equation}

This is an identity. But, if the pure gauge form of $J_{\mu}$ is not 
noted/utilized in the current superfield (only\footnote{That is,
we Did assume  in the current superfield the nontrivial positioning imposed 
on the fermions and the auxiliary fields, which thus satisfy the equation 
identically.}), now dubbed for this purpose {\bf J}, it turns to a constraint 
instead,  with all components vanishing  identically, 
{\em except} for the $\overline{\theta}\theta$ term which consists of (only) 
the zero-curvature constraint for $J_{\mu}$ :
\begin{equation}
\overline{ D}\gamma_p {\bf J} + \overline{\bf J} \gamma_p {\bf J}  
= i~\overline{\theta} \theta~\tau^j\varepsilon^{\mu\nu} (\partial_\mu J^j_\nu +
\varepsilon^{jkl}J^k_\mu J^l_\nu )=0 ~. 
\end{equation}

As a consequence, enforcing this constraint through a Lagrange-multiplier 
superfield ${\bf \Phi}$ 
in an appendage to the 
superspace action eq (\ref{SCMsuperfield}):
\begin{equation}
{\cal L}_{sfield} =
-\frac 14\int d^2 \theta~ \hbox{Tr} \left( ~
\frac 12 \overline{{\bf J}} {\bf J}  +i{\bf \Phi} \cdot \tau ~(
\overline{ D}\gamma_p {\bf J} 
+ \overline{{\bf J}}\gamma_p {\bf J})~\right)  
\end{equation}
will only involve $\Phi^i$ but not $\Psi$ or $Y$, as already observed in 
practice in the component calculation in the previous section. Thus, 
Fridling-Jevicki-type \cite{Fridling} quadrature in superspace will lead 
 to a superfield formulation of the SDSM, below. 

Even though the fermions $\Psi$ appeared to be projected out and thus 
superfluous ``gauge freedom" components of the original ${\bf \Phi}$ 
superfield, they emerge in the final answer below \cite{schonfeld}.
Superspace integration by parts yields the usual quadratic,  
\begin{eqnarray}
&=&-\frac 14 \int d^2 x \int d^2 \theta~ \hbox{Tr}\left(~  
i\overline{\bf J} \gamma_p D {\bf \Phi} \cdot \tau 
+ i{\bf \Phi} \cdot \tau ~\overline{{\bf J}} \gamma_p {\bf J} 
~+\frac 12 \overline{{\bf J}} {\bf J} ~\right) \\
&=&- \frac 14 \int d^2 x \int d^2 \theta~ \left(~  
2i\overline{\bf J}^j \gamma_p D {\bf \Phi}^j 
+  \overline{{\bf J}}^j {\bf M}^{jk}  {\bf J}^k ~\right)   \nonumber \\ 
&=&\frac 14 \int d^2 x \int d^2 \theta~\overline{ D {\bf \Phi}}^j 
{\bf N}^{jk}  D {\bf \Phi}^k
-  (\overline{{\bf J}}^j - i\overline{D {\bf \Phi}}^m \gamma_p {\bf N}^{mj}) 
 {\bf M}^{jk}  ({\bf J}^k +i{\bf N}^{kn} \gamma_p D {\bf \Phi}^n)~,\nonumber
\end{eqnarray}
and a total divergence term discussed below, where 
\begin{equation}
{\bf  M}^{jk}=\delta^{jk}- 2\gamma_p\varepsilon^{jkl}
{\bf \Phi}^l  ~.
\end{equation}
This is the dreibein-twisted-chiral structure observed before, and it is 
inherited by its inverse 
\begin{equation}
{\bf   N}^{jk}=
G_{jk}( {\bf \Phi}) -\gamma_p E_{jk}({\bf\Phi})~,
 \end{equation}
also already encountered in the previous section.
 The customary elimination of 
\begin{equation}
{\bf J}^k= -i~{\bf  N}^{kj}\gamma_p D {\bf \Phi}^j ~ ,
\end{equation} 
identifies, to O$(\theta^0)$, $\chi_j=-\gamma_p (G_{jk} -\gamma_p E_{jk})\Psi^k$
 encountered in eq (\ref{bridge}); and the further terms are
expected to yield the current symphysis formulas.
The algebraic equations for the auxiliary fields result from 
substitution into the superspace action,
\begin{equation}
\int {\cal L}_{sfield} =\frac 14 \int d^2 x \int d^2 \theta~ 
\overline{D}{\bf  \Phi}^j {\bf  N}^{jk} D{\bf \Phi}^k =
\frac 14 \int d^2 x \int d^2 \theta~ 
\overline{D}{\bf\Phi}^j ({\bf G}^{jk}- \gamma_p {\bf E}^{jk}) 
D{\bf \Phi}^k ~.
\end{equation}
This action automatically coincides with ${\cal L}_{sdsm}$, as it is the 
superfield formulation of a $\sigma$-model with torsion possessing 
a given geometry, (cf., e.g., eqn (5.3) of ref \cite{geometrostasis}), 
in this case this new geometry already discussed. 
 
The generating functional $F$ with spinors should emerge out of the total 
divergence  term resulting from the above superspace integration by parts:
\begin{equation}
-\frac 14 ~\int d^2 x \int d^2 \theta~ \hbox{Tr}
\overline{D}\gamma_p \left( 
i{\bf \Phi} \cdot \tau {\bf J } \right) 
=\int d^2 x \int d^2 \theta~ {\overline{\theta}\theta\over 2} ~\partial^\nu 
\varepsilon_{\nu\mu} (J^{\mu~j} \Phi^j 
+{i\over2} \bar\chi^j\gamma^\mu \Psi^j 
-{i\over2} \bar\chi^j\gamma^\mu \varepsilon^{jkl}\Phi^k \chi^l),
\end{equation}
i.e. 
\begin{equation}
=\int\! dx ~(J^{1~j} \Phi^j+{i\over2} \bar\chi^j\gamma^1 \Psi^j 
-{i\over2} \bar\chi^j\gamma^1 \varepsilon^{jkl}\Phi^k \chi^l).
\end{equation}
However, superfield formulations of general $\sigma$-models
with torsion always contain a total divergence of a fermion bilinear term in 
the reduction from superspace, cf.~eqn (5.8) in \cite{geometrostasis}.
In this case, this extra term contributes 
$-{i\over2}\overline{\Psi}^j\gamma^1\varepsilon^{jkl}\Phi^k\Psi^l/(1+4\Phi^2)$
 to the above, to produce the generating functional 
\begin{equation}
F'=\int\! dx ~\left(J^{1~j} \Phi^j+{i\over2} \bar\chi^j\gamma^1 \Psi^j 
-{i\over2} \bar\chi^j\gamma^1 \varepsilon^{jkl}\Phi^k \chi^l
-{i\over2}\overline{\Psi}^j\gamma^1\varepsilon^{jkl}{\Phi^k\over (1+4\Phi^2)}
\Psi^l  \right).
\end{equation}

Variation with respect to $\bar \chi$ yields $\bar\pi_{\chi}$, hence
\begin{equation}
\gamma_p \chi^j= \Psi^j -2\varepsilon^{jkl}\Phi^k \chi^l~, 
\end{equation}
so that 
\begin{equation}
\gamma_p \chi^j= (V_{(aj)} +\gamma_p V_{[aj]} )\Psi^a ~, 
\end{equation}
the right-hand-side fermion actually being $X$ as introduced in 
(\ref{bridge}). Note that we never really introduced tangent space fermions 
for the dual theory in superspace---they emerge as an output. 
Moreover, variation with respect to $\overline{\Psi}$ yields $-\bar\pi_{\Psi}$
 for  ${\cal L}_{sdsm}$, hence
\begin{equation}
\gamma_p G_{jk} \Psi^k= \chi^j +E_{jk} \Psi^k,
\end{equation}
and hence 
\begin{equation}
\gamma_p \chi^j= G_{jk} \Psi^k -  \gamma_p E_{jk} \Psi^k = X^j,
\end{equation}
consistently to the above. 

Eliminating $\Psi's$ in 
favor of $X$'s in $F'$ yields 
\begin{equation}
F'=\int\! dx ~\left(J^{1~j} \Phi^j
+{i\over2} \bar\chi^j\gamma^1 X^j 
-{i\over2} (\bar\chi^j+ \bar X^j\gamma^p)\gamma^1 \varepsilon^{jkl}\Phi^k 
(\chi^l- \gamma^p X^l) \right).
\end{equation}
This provides an alternate route to the same theory, and the final term 
vanishes given the above results of the transformation. 

\noindent \rule{7in}{0.1em}

\section*{Acknowledgments}
 
Work supported by the NSF grant PHY-95-07829 and
the U.S.~Department of Energy, Division of High Energy Physics, Contract
W-31-109-ENG-38. T. U. and T. C. thank Argonne National Laboratory for its 
hospitality during important phases of this project. 
We are thankful to  L. Palla for communicating to us a draft of \cite{palla} 
before release. 

\noindent \rule{7in}{0.1em}

\renewcommand{\theequation}{\Alph{section}.\arabic{equation}}
\appendix
\section{Appendix: Overview of Canonical Transformations in Mechanics}  
\setcounter{equation}{0}
We review some rarely emphasized aspects of canonical transformations in 
classical mechanics: we take as our starting point invariance of Poisson 
Brackets (PB), instead of the more conventional preservation of Hamilton's 
equations. Poisson Brackets are suitable for eventual quantization, 
$\{u,v  \}\mapsto { 1\over i\hbar} [u,v]$, as they turn into canonical 
commutators. Extension to field theory merely involves arraying a continuum of 
modes, and transcription of sums into integrals, as exemplified in the text.

Poisson Brackets can be defined as
\begin{equation} 
\{u,v \}_{qp}\equiv {\partial   u \over \partial q  }
{\partial v    \over \partial p }-
{\partial u    \over \partial p  }
{\partial v   \over \partial q }.
\end{equation}

PB's are antisymmetric; linear; they obey the Jacobi identity (from 
associativity of the underlying operators);
and convert by the chain rule:
\begin{equation} \{u,v\}_{QP}=\{u,v\}_{qp}\{q,p\}_{QP}.
\end{equation}
Now the transformations 
\begin{equation}
(q,p) \mapsto (Q(q,p), P(q,p)) \end{equation}
are called {\em canonical} when they yield trivial Jacobians,
\begin{equation}
\{  Q, P  \}_{qp} = 1  \qquad\qquad  \hbox{hence} \qquad\qquad
\{  q, p  \}_{QP} = 1,
 \end{equation}
so that they preserve the Poisson Brackets (``canonical invariants") of their 
functions. Equivalently, 
\begin{equation}
\{  q, p  \}=\{  Q, P  \},
 \end{equation}
in any basis. Following Poincar\'{e}, the measure of phase-space area/volume is 
seen to be preserved,
\begin{equation}
dQdP= dq dp ~\{  Q, P  \}  .
 \end{equation}

For instance, point transformations (which generalize to usual local field 
redefinitions in field theory) are
\begin{equation}
Q=J(q), \qquad P=p/J'(q) \qquad\qquad\Longrightarrow \qquad
\{  J(q), ~p/J'(q)  \}_{qp}=1. 
 \end{equation}

Now consider the transformation {\em generated} by the hybrid function 
$F_2(q,P)$:
\begin{equation}
p={\partial F_2(q,P) \over \partial q } , \qquad \qquad 
Q={\partial F_2(q,P) \over \partial P },
 \end{equation}
where, in principle, the first equation can be inverted to produce 
$P(q,p)$, which is then substituted into the 2nd to yield $Q$ as a function of 
$q,p$. 
(The point transformation just illustrated is generated by $F_2=PJ(q)$.)
This transformation is canonical, seen explicitly as follows.

Define  $p=\partial F_2(q,P)/\partial q|_P=~'\!F$  and 
$Q=\partial F_2(q,P)/\partial P|_q=F'$, contradistinguishing the arguments 
being differentiated. Now switch basis to $q,p$, and 
consider $P$ as a function of $q,p$, partly specified via the partial 
differential equation $\partial p/ \partial p=1 =~ '\!F' P_{,p}$.
It is then straightforward to show in this $q,p$ basis that 
\begin{equation} 
1=\{q,~ '\!F\} = \{ F',P\}, \end{equation}
since the middle expression equals 1 of the l.h.s. 
by the above differential equation, and, by the same token, the r.h.s.~also 
equals 
\begin{equation}
( F'' P_{,q} +~ '\!F') P_{,p}  - F'' P_{,p} P_{,q}=1.
\end{equation}

Infinitesimally, this transformation is also easily seen to be canonical, as 
follows.
Introduce a generating function infinitesimally expanded around the identity 
through an expansion parameter $w$: 
\begin{equation} 
   F_2(q,P)=qP - w G(q,P). 
\end{equation}
The $O(w^0)$ piece is the identity, as
\begin{equation}
p=P -w {\partial G(q,P) \over \partial q } , \qquad \qquad \qquad 
Q=q-w {\partial G(q,P) \over \partial P } .
\end{equation}
To leading order in $w$, one can substitute $G(q,p)$ for $G(q,P)$,
so that 
\begin{equation} 
Q-q=-w {\partial G(q,p) \over \partial p }+O(w^2)=w\{G,q\}_{qp}+O(w^2) ,
\end{equation}
\begin{equation} 
P-p=w {\partial G(q,p) \over \partial q }+O(w^2)=w\{G,p\}_{qp} +O(w^2). 
\end{equation}
Then it is easy to see, by the Jacobi identity, that this transformation 
is canonical to $O(w^2)$:
\begin{equation} 
           \{q+w\{G,q\},~ p+w\{G,p\}\}=1+w \{G,\{q,p\}\}=1.
\end{equation}

Actually, the full, exponentiated transformation 
\begin{equation} 
Q= e^{w\{G} q \equiv  q +w \{G,q\} + w^2 \{G,\{G,q\}\}/2!+...
\end{equation}   
\begin{equation} 
P= e^{w\{G} p \equiv  p +w \{G,p\} + w^2 \{G,\{G,p\}\}/2!+... 
\end{equation}   
is canonical to all-orders in $w$, 
\begin{equation} 
Y(w)\equiv \{Q,P\} =1 ,
\end{equation}
for essentially the same reason. \newline
{\bf Proof:} Note that, from the very definition of the Hadamard exponential 
operator above, for any $w-$independent $a$, such as $q$ or $p$,
\begin{equation} 
{d~(e^{w\{G} a) \over dw}   = \{G,~ e^{w\{G} a \} .
\end{equation} 
Consequently, by the Jacobi identity:
\begin{equation} 
{dY(w)\over dw}  = \{\{G,Q\}, P\} + \{Q, \{G,P\}\}= \{G,Y(w)\}.
\end{equation}
Now $Y(w)= 1 + \sum_{n=1}^{\infty}  w^n y_n$ 
yields the recursion 
\begin{equation} 
y_{n+1}=\{G,y_n\}, 
\end{equation}
and all the $y_{n>0}=0$ by induction based on $y_0=1$.
Consequently $Y(w) =1$.\qquad  \rule{.5\baselineskip}{.5\baselineskip}\newline 
Note this is a standard integrated Lie-evolution, but it is still hard 
to fully connect to the unfolded general $F_2$.

The other three types of canonical transformation can be 
obtained through the trivial symplectic reflection, which is  
is also canonical:
\begin{equation} 
(q,p) \mapsto (Q=-p,P=q)  \qquad \hbox{since} \qquad \{-p,q\}=1,
\end{equation}
and thus may combine with other canonical transformations to yield more 
canonical transformations, by the chain rule of PBs. 
Applied to the variables $Q,P$ of the above $F_2$, it produces the canonical 
transformation, generated by the {\em same} functional form, now called $F_1$:
\begin{equation}
p={\partial F_1(q,Q) \over \partial q } , \qquad \qquad \qquad 
P=-{\partial F_1(q,Q) \over \partial Q },
 \end{equation}
and likewise for the other two types.

The Field Theoretical construction in this article on the $\sigma$-model 
is of the $F_1(q,Q)$ type, but it could be trivially converted into the 
$F_2(q,P)$ type by this trivial symplectic reflection (readily generated 
by $F_1(q,Q)= qQ$).
The classical mechanics analog of our transformation is $F_1(q,Q)=Q~ J(q)$, 
which then goes by the above symplectic reflection to 
\begin{equation}
F_2(q,P)= -PJ(q), 
 \end{equation}
s.t. 
\begin{equation} 
- {\partial F_2(q,P) \over   \partial P} = Q, \qquad \qquad \qquad 
-{ \partial F_2(q,P)\over   \partial q} = p ,
  \end{equation}                                           
thus a point transformation:
\begin{equation}   Q= J(q) \end{equation}
\begin{equation}   
p= P {\partial J(q) \over  \partial q} , ~~~~\hbox{hence}~~~~~~~
P={p \over \partial J(q) / \partial q}. 
\end{equation}
In other words, the canonical transformation used, in more conventional 
$F_2$ 
language, is simply a transition from the {$q$'s to the $J(q)$'s with the 
standard point determinant scaling for the momenta to preserve the PB's ,
\begin{equation}   
\{ J(q), p {1 \over \partial J(q) / \partial q} \}=1. 
\end{equation}
In field theory, it is a transition from $\varphi$ to $J_1(\varphi)$. But
interchange of $\Phi$'s and $\Pi$'s in the Hamiltonian of the DSM,
yields something unconventional, involving space derivatives of the 
$\Pi$s. 

Now, how is the identity transformation generated by $F_1(q,Q)$, 
and why should one choose to base the discussion on $F_2$, in the first place?  
Some awkward features of the $F_1(q,Q)$ generating function have been pointed 
out by Schwinger \cite{Schwinger}. The inverse canonical
transformation is, evidently, 
\begin{equation}
F_1(Q,q)=- F_1(q,Q).  
 \end{equation}
The composition of two successive transformations 
$(q,p)\mapsto (\bar{q},\bar{p})\mapsto(Q,P)$ is simply generated by the 
the {\em sum} of the respective generating functions, 
\begin{equation}
W(q,Q)=F_1(q,\bar{q})+F_1(\bar{q},Q),
\end{equation}
where each term generates the respective piece of the total transformation,
and the dependence on the intermediate point (``superfluous variable") 
vanishes between the two in the total transformation,
\begin{equation}
{\partial W(q,Q) \over \partial \bar{q}}=0 .
 \end{equation}
But, by the above defined inverse, the identity should be then generated by 
$W(q,Q)=0$. This singularity of the generator can be made more palatable by 
retaining the superfluous variable $\bar{q}$ which serves as a Lagrange 
multiplier:  
 \begin{equation} 
W(q,Q)=\bar{q} (q-Q) ,
\end{equation}
hence 
\begin{equation} 
{\partial W\over \partial \bar{q} }= 0 \qquad\Longrightarrow \qquad 
q=Q,
\end{equation}
\begin{equation}
{\partial W\over \partial q }= \bar{q} = -{\partial W\over \partial Q }
 \qquad\Longrightarrow \qquad 
p=P.
 \end{equation}

{\em Motion is a canonical transformation}, of the above 
infinitesimal $F_2$ type. When $G$ is chosen to be the hamiltonian $H$, 
and $w=-dt$,  
\begin{equation}
dq=dt~ \{q,H  \}_{qp}  ,\qquad \qquad dp=dt~ \{p,H  \}_{qp}  ,
\end{equation}
which comprise Hamilton's Equations,
\begin{equation}
\dot{q}=\partial H /\partial p ,\qquad \qquad \dot{p}=-\partial H /\partial q, 
\end{equation}
dictating incompressible flow of the phase fluid, 
$\partial \dot{q}  /\partial q + \partial \dot{p} /\partial p=0$. 
(As seen above, such a transformation readily exponentiates.) 
Consequently, for any function $f(q,p)$,
\begin{equation}
\{ f(q,p), H\}_{qp}=   { df\over dt}. 
\end{equation}
Now consider some arbitrary canonical transformation to $Q,P$, and take $f$ to 
be $Q$ and $P$, respectively; switching PB basis by virtue of the canonical 
nature of the transformation, one sees directly that 
\begin{equation} 
\dot{Q}=\{ Q, H\}_{QP}=\partial H /\partial P ,\qquad \qquad 
\dot{P}=\{ P, H\}_{QP}=-\partial H /\partial Q. 
\end{equation}
That is, any canonical transformation also preserves Hamilton's Equations 
of motion:
\def\mapright#1{\vbox{\ialign{##\crcr $\hfil\scriptstyle{\ #1 \ }\hfil$   
    \crcr\noalign{\kern+1pt\nointerlineskip}\rightarrowfill \crcr} }}
\def\mapdownl#1{\lower1.4ex\hbox{\llap{$\vcenter{\hbox{$\scriptstyle#1$}}$}}
     \lower1.4ex\hbox{\Big\downarrow}}
\def\mapdownr#1{\lower1.4ex\hbox{\Big\downarrow} \lower1.4ex\hbox{
    \rlap{$\vcenter{\hbox{$\scriptstyle#1$}}$}}}
\begin{equation} 
\matrix{    q_t  & \mapright{~~~H~~~} & q_T \cr 
\mapdownl{F} & \  & \mapdownr{F} \cr
Q_t & \mapright{~~~H~~~} & Q_T.  \cr                }
\end{equation}

Motion is also generated by the Action Integral $S$ (Hamilton's principal 
function), but {\em on the classical path}, via an $F_1$ transformation; 
this is the one utilized by Dirac in his celebrated quantum Hamilton-Jacobi 
functional integral \cite{Dirac33}: 
\begin{equation}
p_t = {\delta \int_T^t d\tau L \over \delta q_t}  \qquad 
p_T =- {\delta \int_T^t d\tau L \over \delta q_T}~.
 \end{equation}
The intermediate-times variables $q(t)$ are {\em not arbitrary}, but must 
be specified by the equations of motion. To appreciate this, recall that 
the action integral  may be effectively 
regarded as a sum of infinitesimal transformation generators of type $F_1$, 
namely $dt~ L({q+Q\over 2},{q-Q\over dt})$; thus 
\begin{equation}
 \delta \int_T^t d\tau L = 
\int_T^t d\tau ~ \delta q  ( {\delta L \over \delta q}
 -{d\over dt}  {\delta L \over \delta \dot{q}})
 +  {\delta L_t \over \delta \dot{q}_t}  \delta q_t  -  
  {\delta L_T \over \delta \dot{q}_T} \delta q_T ~ .\end{equation}
Vanishing of the first term in parenthesis (the Euler-Lagrange equations of 
motion) to yield the above result is dictated by the requirement of 
independence from the (intermediate times) superfluous variables. 
That is, {\bf the  requirement of
continuous canonical transformation for motion underlies the classical 
variational principle.}
 Dirac \cite{Dirac33} discovered that in Quantum 
Mechanics the generator must be exponentiated and the superfluous variables 
must be integrated over instead---the above classical path is then only the 
contribution to leading order in $\hbar$.

\section{Appendix: Explicit Geometry of the Dual Models}
\setcounter{equation}{0} The dual sigma model is not a WZWN model on a group
manifold. The geometry of the DSM is described in detail by the following,
in the conventions of ref \cite{geometrostasis}.
\begin{equation}
G_{ab}=V_a^{\;j}V_b^{\;j}=\frac 1{1+4\Phi ^2}\left( \delta _{ab}+4\Phi
^a\Phi ^b\right) \,,
\end{equation}
\begin{equation}
\det G={1/(1+4\Phi ^2)^2},
\end{equation}
\begin{equation}
E_{ab}=\frac 1{1+4\Phi ^2}\left( -2\varepsilon ^{abc}\Phi ^c\right) ,
\end{equation}
\begin{equation}
V_a^{\;j}=G_{aj}+E_{aj}=\frac 1{1+4\Phi ^2}\left( \delta _{aj}+4\Phi ^a\Phi
^j-2\varepsilon ^{ajc}\Phi ^c\right) ,
\end{equation}
\begin{equation}
\det V_a^j=\sqrt{\det G}=(1+4\Phi ^2)^{-1}.
\end{equation}
\begin{equation}
V^{aj}=\delta _{aj}-2\varepsilon ^{ajc}\Phi ^c,
\end{equation}
\begin{equation}
G_{ab}^{-1}=V^{aj}V^{bj}=\left( 1+4\Phi ^2\right) \delta _{ab}-4\Phi ^a\Phi
^b.
\end{equation}
Note that, in this remarkable geometry, base- and target-space
indices are innocuously interchangeable {\em for our choice of coordinates} (%
$\Phi $'s), since $\Phi _a=G_{ab}\Phi ^b=\Phi ^a$, $\Phi ^h=\Phi ^aV_a^{~h}$.

Connections are obtained in the usual way:
\begin{eqnarray}
\Gamma _{abc} &=&\frac 12\left( \partial _bG_{ac}+\partial _cG_{ab}-\partial
_aG_{bc}\right)  \nonumber \\
\ &=&\frac 4{1+4\Phi ^2}\left( \Phi ^a\delta _{bc}+\Phi ^aG_{bc}-\Phi
^bG_{ac}-\Phi ^cG_{ab}\right),
\end{eqnarray}
\begin{eqnarray}
\Gamma _{\;bc}^a &=&\frac 4{1+4\Phi ^2}\left( \Phi ^a\delta _{bc}+\Phi
^aG_{bc}-\Phi ^b\delta _{ac}-\Phi ^c\delta _{ab}\right)  \nonumber \\
&=&16{\frac{\Phi ^a\Phi ^b\Phi ^c}{(1+4\Phi ^2)^2}}+8\Phi ^a\delta ^{bc}{%
\frac{(1+2\Phi ^2)}{(1+4\Phi ^2)^2}}-{\frac{4\Phi ^b\delta ^{ac}+4\Phi
^c\delta ^{ab}}{(1+4\Phi ^2)}}~,
\end{eqnarray}
\begin{eqnarray}
S_{abc} &=&\frac 12\left( \partial _aE_{bc}+\partial _bE_{ca}+\partial
_cE_{ab}\right)  \nonumber \\
&=&\frac{(3+4\Phi ^2)}{\left( 1+4\Phi ^2\right) ^2}\,\left( -\varepsilon
^{abc}\right).
\end{eqnarray}
Note that the DSM dreibein does\ not satisfy Cartan-Maurer equations.
Rather,
\begin{equation}
\partial _aV_b^{\;j}-\partial _bV_a^{\;j}+4\varepsilon
^{jkl}V_a^{\;k}V_b^{\;l}=-4\frac{\left( 1-4\Phi ^2\right) }{\left( 1+4\Phi
^2\right) ^2}\,\left( \Phi ^a\delta _{bj}-\Phi ^b\delta _{aj}\right) -\frac{%
16}{\left( 1+4\Phi ^2\right) ^2}\,\left( \Phi ^b\varepsilon ^{ajc}\Phi
^c-\Phi ^a\varepsilon ^{bjc}\Phi ^c\right),
\end{equation}
where
\begin{equation}
\varepsilon ^{jkl}V_a^{\;k}V_b^{\;l}=\frac{-2}{1+4\Phi ^2}\,\left( \Phi
^a\delta _{bj}-\Phi ^b\delta _{aj}\right) +\frac 1{1+4\Phi ^2}\,\varepsilon
^{abj}.
\end{equation}
Alternatively,
\begin{equation}
\partial _aV_b^{\;j}-\partial _bV_a^{\;j}=\frac{-4}{1+4\Phi ^2}\,\varepsilon
^{abc}G_{cj}-4\varepsilon ^{abc}S_{cjd}\Phi ^d.
\end{equation}
Contrast these to the corresponding relations for the CM,
\begin{equation}
\partial _av_b^{\;j}-\partial _bv_a^{\;j} =2\varepsilon
^{jkl}v_a^{\;k}v_b^{\;l} = 2\varepsilon ^{abj}+\frac{-2}{\sqrt{1-\varphi ^2}}%
\,\left( \varphi ^a\delta _{bj}-\varphi ^b\delta _{aj}\right) .
\end{equation}

However, note that, in contrast to group manifolds,
\begin{equation}
S_{abc}=-{\frac{3+4\Phi ^2}{1+4\Phi ^2}}~\varepsilon
^{jkl}V_a^{~j}V_b^{~k}V_c^{~l}~.
\end{equation}
So, e.g.,
\begin{equation}
\partial _aG_{bj}-\partial _bG_{aj}=2S_{abc}V^{[cj]}.
\end{equation}
Hence
\begin{equation}
\Gamma _{[ab]c}=S_{abd}V^{[dc]}.
\end{equation}

The geometry follows from direct, albeit lengthy computation 
(using Maple, http://daisy.waterloo.edu/): 
\begin{equation}
D_dS_{abc}={\frac{16}{(1+4\Phi ^2)^3}\;}\Phi ^d\;\varepsilon _{abc}\;,
\end{equation}
\begin{equation}
R_{\;bcd}^a={\frac 4{(1+4\Phi ^2)^3}\;}\left(
\begin{array}{c}
\left( 3+12\Phi ^2+16\Phi ^4\right) \;\left( \delta ^{ac}\delta ^{bd}-\delta
^{ad}\delta ^{bc}\right) \\
+\left( 16\Phi ^2\right) \;\Phi ^b\left( \Phi ^c\delta ^{ad}-\Phi ^d\delta
^{ac}\right) \\
+\left( 12+16\Phi ^2\right) \;\Phi ^a\left( \Phi ^d\delta ^{bc}-\Phi
^c\delta ^{bd}\right)
\end{array}
\right) ,
\end{equation}
\begin{equation}
S_{fad}S_{\;cb}^f-S_{fac}S_{\;db}^f={\frac{\left( 3+4\Phi ^2\right) ^2}{%
(1+4\Phi ^2)^4}\;}\left(
\begin{array}{c}
\;\left( \delta ^{ac}\delta ^{bd}-\delta ^{ad}\delta ^{bc}\right) \\
-4\Phi ^b\left( \Phi ^c\delta ^{ad}-\Phi ^d\delta ^{ac}\right) \\
-4\Phi ^a\left( \Phi ^d\delta ^{bc}-\Phi ^c\delta ^{bd}\right)
\end{array}
\right) \;,
\end{equation}
\begin{equation}
S_{\;df}^aS_{\;cb}^f-S_{\;cf}^aS_{\;db}^f={\frac{\left( 3+4\Phi ^2\right) ^2%
}{(1+4\Phi ^2)^3}\;}\left(
\begin{array}{c}
\;\left( \delta ^{ac}\delta ^{bd}-\delta ^{ad}\delta ^{bc}\right) \\
-4\Phi ^b\left( \Phi ^c\delta ^{ad}-\Phi ^d\delta ^{ac}\right)
\end{array}
\right) \;  .
\end{equation}
Thus, the torsionful Riemann tensor amounts to
\begin{eqnarray}
{\cal R}_{\;bcd}^a &\equiv
&R_{\;bcd}^a-S_{\;df}^aS_{\;cb}^f+S_{\;cf}^aS_{\;db}^f+D_cS_{\;bd}^a-D_dS_{%
\;bc}^a  \nonumber \\
&=&{\frac 3{(1+4\Phi ^2)}\;}\left( \delta ^{ac}\delta ^{bd}-\delta
^{ad}\delta ^{bc}\right) \\
&&+{\;\frac{4\,\left( 9+4\Phi ^2\right) }{(1+4\Phi ^2)^2}\;}\Phi ^b\left(
\Phi ^c\delta ^{ad}-\Phi ^d\delta ^{ac}\right) +{\frac{16\,\left( 3+4\Phi
^2\right) }{(1+4\Phi ^2)^3}\;}\Phi ^a\left( \Phi ^d\delta ^{bc}-\Phi
^c\delta ^{bd}\right)  \nonumber \\
&&+{\frac{16}{(1+4\Phi ^2)^2}\;}\left( \Phi ^c\varepsilon _{abd}-\Phi
^d\varepsilon _{abc}\right) \;+{\frac{64}{(1+4\Phi ^2)^3}\;}\Phi ^a\Phi
^f\left( \Phi ^d\varepsilon _{bcf}-\Phi ^c\varepsilon _{bdf}\right) \;.
\end{eqnarray}
Actually, curly-R without a raised index is a little easier to look at,
since it has the usual antisymmetries under interchange of pairs of indices,
but not the symmetry under interchange of pairs of pairs that the
torsionless curvatures have.
\begin{eqnarray}
{\cal R}_{abcd} &=&{\frac 3{(1+4\Phi ^2)^2}\,}\left( \delta ^{ac}\delta
^{bd}-\delta ^{ad}\delta ^{bc}\right)  \nonumber \\
&&+{\frac{4\,(9+4\Phi ^2)}{(1+4\Phi ^2)^3}\,}\left( \Phi ^a\left( \Phi
^d\delta ^{bc}-\Phi ^c\delta ^{bd}\right) -\Phi ^b\left( \Phi ^d\delta
^{ac}-\Phi ^c\delta ^{ad}\right) \right) {\;}  \nonumber \\
&&+{\frac{16}{(1+4\Phi ^2)^3}\,}\left( \Phi ^c\varepsilon _{abd}-\Phi
^d\varepsilon _{abc}\right) \;.
\end{eqnarray}

There is an alternate route to this torsionful curvature. 
The spin connection \cite{geometrostasis} is worked out to be 
\begin{eqnarray}
\Omega _a^{\;ij} &\equiv &V^{bi}{\cal D}_aV_b^{\;j}\equiv V^{bi}\left(
D_aV_b^{\;j}+S_{abc}V^{cj}\right)  \nonumber \\
&=&\frac 1{\left( 1+4\Phi ^2\right) ^2}\,\left( (14+24\Phi ^2)\left( \delta
^{aj}\Phi ^i-\delta ^{ai}\Phi ^j\right) +(-5-4\Phi ^2)\left( \varepsilon
^{aij}+4\Phi ^a\varepsilon ^{ijm}\Phi ^m\right) \right)  \nonumber \\
&=&\frac{(14+24\Phi ^2)}{\left( 1+4\Phi ^2\right) ^2}\,\left( \delta
^{aj}\Phi ^i-\delta ^{ai}\Phi ^j\right) -\frac{(5+4\Phi ^2)}{\left( 1+4\Phi
^2\right) ^2}\,\left( \varepsilon ^{aij}+4\Phi ^a\varepsilon ^{ijm}\Phi
^m\right) \;.
\end{eqnarray}
This is not the spin connection on any group manifold. On the one hand,
first differentiate the spin connection and then combine it with its square,
antisymmetrically, to yield the curly-curvature:
\begin{equation}
\!\!\partial _a\Omega _b^{\;ij}-\partial _b\Omega _a^{\;ij}+\Omega
_a^{\;ik}\Omega _b^{\;kj}-\Omega _b^{\;ik}\Omega _a^{\;kj}=\frac 1{\left(
1+4\Phi ^2\right) ^3}\,\left(\!\!\!\!
\begin{array}{c}
3\times \left( 1+4\Phi ^2\right) ^2\left( \delta ^{bj}\delta ^{ai}-\delta
^{bi}\delta ^{aj}\right) \\
-16\!\times\!\left( 4\Phi ^2+5\right) \left( \Phi ^a\left( \delta ^{bj}\Phi
^i-\delta ^{bi}\Phi ^j\right) -\Phi ^b\!\left( \delta ^{aj}\Phi ^i-\delta
^{ai}\Phi ^j\right) \right) \\
-2\times \left( 16\Phi ^4+24\Phi ^2-11\right) \left( \Phi ^a\varepsilon
^{bij}-\Phi ^b\varepsilon ^{aij}\right)
\end{array}\!\!\!
\right)\!.
\end{equation}

On the other hand, for our dreibein
\begin{equation}
({\cal R}^{ij})_{ab}\equiv V^{ci}V^{dj}{\cal R}_{cdab}=\left( 1+4\Phi
^2\right) {\cal R}_{ijab}+4\left( \Phi ^i{\cal R}_{jmab}-\Phi ^j{\cal R}%
_{imab}\right) \Phi ^m+\left( \varepsilon ^{cdi}\Phi ^j-\varepsilon
^{cdj}\Phi ^i\right) {\cal R}_{cdab}\;,
\end{equation}
which gives explicit agreement with the RHS of the previous expression
(providing some algebraic checks) so
\begin{eqnarray}
({\cal R}^{ij})_{ab}=\partial _a\Omega _b^{\;ij}-\partial _b\Omega
_a^{\;ij}+\Omega _a^{\;ik}\Omega _b^{\;kj}-\Omega _b^{\;ik}\Omega _a^{\;kj},
\end{eqnarray}
as it should be.

Recall, for three-dimensional manifolds, the Weyl tensor vanishes (even when
torsion is present, as in the model at hand), permitting us to re-express
the Riemann tensor in terms of Ricci and scalar curvatures:
\begin{equation}
{\cal R}_{abcd}=\frac 12(G_{ad}G_{bc}-G_{ac}G_{bd}){\cal R}+G_{ac}{\cal R}%
_{bd}-G_{ad}{\cal R}_{bc}+G_{bd}{\cal R}_{ac}-G_{bc}{\cal R}_{ad} \qquad\qquad
\end{equation}
\[           \qquad
=\left( {\cal R}_{ac}-\frac 14\;G_{ac}{\cal R}\right) G_{bd}-\left( {\cal R}%
_{ad}-\frac 14\;G_{ad}{\cal R}\right) G_{bc}+G_{ac}\left( {\cal R}%
_{bd}-\frac 14\;G_{bd}{\cal R}\right) -~G_{ad}\left( {\cal R}_{bc}-\frac
14\;G_{bc}{\cal R}\right) \;.
\]
(N.B. $a,b,c=1,2,3$ only.) On the other hand, for three-dimensional
manifolds, the issue of conformal flatness is decided not by the Weyl
tensor, but rather by the Cotton tensor \cite{cotton}, obtained by taking
derivatives of the Ricci and scalar curvature combinations exhibited in the
last equality. Without torsion, the Cotton tensor is defined as
\begin{equation}
C_{abc}=D_c\left( R_{ab}-\frac 14\;G_{ab}R\right) -D_b\left( R_{ac}-\frac
14\;G_{ac}R\right) .
\end{equation}
The manifold is conformally equivalent to a flat space iff $C_{abc}=0.$ It
is straightforward to check that $C_{abc}=0$ for the dual sigma model. (This
is also true, rather obviously, for the usual chiral model.)

Along these lines, it is interesting to note the same linear combination of
Ricci and scalar curvature appears in the quartic spinor terms for a general
supersymmetric model defined on a three-dimensional manifold. Taking into
account the Majorana property of the spinors, and making Fierz
rearrangements, gives
\begin{eqnarray}
{\cal R}_{abcd}\overline{\Psi }^a(1+\gamma _p)\Psi ^c\overline{\Psi }%
^b(1+\gamma _p)\Psi ^d &=&4G_{ab}(\overline{\Psi }^a\Psi ^b)\left( {\cal R}%
_{cd}-\frac 14\;G_{cd}{\cal R}\right) \overline{\Psi }^c(1+\gamma _p)\Psi ^d
\\
&=&4G_{ab}(\overline{\Psi }^a\Psi ^b)\left( {\cal R}_{(cd)}-\frac 14\;G_{cd}%
{\cal R}\right) \overline{\Psi }^c\Psi ^d+4G_{ab}(\overline{\Psi }^a\Psi
^b)\;{\cal R}_{[cd]}\overline{\Psi }^c\gamma _p\Psi ^d\; .  \nonumber
\end{eqnarray}
(N.B. $a,b,c=1,2,3$ only.)

Another way to say this, valid for higher dimensional manifolds, is to write
the quartic term in general cases using the Weyl tensor:
\begin{equation}
{\cal C}_{abcd}{\cal \ =R}_{abcd}+\frac 12(G_{ac}G_{bd}-G_{ad}G_{bc}){\cal R}%
-G_{ac}{\cal R}_{bd}+G_{ad}{\cal R}_{bc}-G_{bd}{\cal R}_{ac}+G_{bc}{\cal R}%
_{ad}~.
\end{equation}
\begin{equation}
{\cal R}_{abcd}\overline{\Psi }^a(1+\gamma _p)\Psi ^c\overline{\Psi }%
^b(1+\gamma _p)\Psi ^d=
\end{equation}
$$
={\cal C}_{abcd}\overline{\Psi }^a(1+\gamma _p)\Psi ^c%
\overline{\Psi }^b(1+\gamma _p)\Psi ^d+4G_{ab}(\overline{\Psi }^a\Psi
^b)\left( {\cal R}_{cd}-\frac 14\;G_{cd}{\cal R}\right) \overline{\Psi }%
^c(1+\gamma _p)\Psi ^d.
$$

Specification of the curly curvature now allows evaluation of the curly-Ricci
tensor, crucial to the computation of the $\beta$-function to one loop 
\cite{geometrostasis}: 
\begin{equation}
{\cal R}_{(bd)}={\frac 2{(1+4\Phi ^2)^3}\;}\left( (3+16\Phi ^4)\;\delta
^{bd}-4\;(3+32\Phi ^2+16\Phi ^4)\Phi ^b\Phi ^d\right) \;.
\end{equation}
There is also the antisymmetric part of the curly-Ricci tensor,
\begin{equation}
{\cal R}_{[bd]}=D_aS_{\;bd}^a={\frac{16}{(1+4\Phi ^2)^3}\;}\varepsilon
^{bdf}\Phi ^f=-8\,\left( \det G\right) \;E_{ab}\;.
\end{equation}
(We disagree here with the corresponding expressions of ref \cite{subbotin}, 
and agree with \cite{palla}; we are grateful to L. Palla for
communicating to us this result before publication). Finally, there is the
scalar curly-curvature:
\begin{equation}
{\cal R}={\frac 2{(1+4\Phi ^2)^2}\;}(3-4\Phi ^2)^2\;=2\,\left( {\frac
4{1+4\Phi ^2}-1}\right) ^2,
\end{equation}
with small and large field limits, 18 and 2, respectively. I.e. the
curvature starts as a sphere of $18/r^2$, goes to zero at $\Phi ^2=3/4$, and
stabilizes asymptotically to $2/r^2$ . 

It goes without saying that full appreciation of this new geometry should 
hold the key to renormalization of the dual models beyond one loop.

\noindent \rule{7in}{0.1em}

\end{document}